# Critical Current Density and Resistivity of MgB$_2$ Films

J. M. Rowell[a]

*Department of Chemical and Materials Engineering,*
*Arizona State University, Tempe, AZ 85287*

S. Y. Xu, X. H. Zeng[b], A. V. Pogrebnyakov

*Department of Physics, The Pennsylvania State University, University Park, PA 16802*

Qi Li

*Department of Physics and Materials Research Institute,*
*The Pennsylvania State University, University Park, PA 16802*

X. X. Xi

*Department of Physics, Department of Materials Science and Engineering, and*
*Materials Research Institute, The Pennsylvania State University, University Park, PA 16802*

J. M. Redwing

*Department of Materials Science and Engineering and Materials Research Institute,*
*The Pennsylvania State University, University Park, PA 16802*

W. Tian and Xiaoqing Pan

*Department of Materials Science and Engineering,*
*The University of Michigan, Ann Arbor, MI 48109*

## Abstract

The high resistivity of many bulk and film samples of MgB$_2$ is most readily explained by the suggestion that only a fraction of the cross-sectional area of the samples is effectively carrying current. Hence the supercurrent ($J_c$) in such samples will be limited by the same area factor, arising for example from porosity or from insulating oxides present at the grain boundaries. We suggest that a correlation should exist, $J_c \sim 1/\Delta\rho_{300-50K}$, where $\Delta\rho_{300-50K}$ is the change in the apparent resistivity from 300 K to 50 K. We report measurements of $\rho(T)$ and $J_c$ for a number of films made by hybrid physical-chemical vapor deposition which demonstrate this correlation, although the "reduced effective area" argument alone is not sufficient. We suggest that this argument can also apply to many polycrystalline bulk and wire samples of MgB$_2$.

The resistivity of MgB$_2$, as reported in single crystal, polycrystalline bulk, thin film and wire samples, varies by orders of magnitude. In single crystals, the resistivity is reported by Yu. Eltsev *et al*. [1, 2] to be 5.3 µΩ.cm at 300 K, and 1.0 µΩ.cm at 50 K. Some polycrystalline bulk and thin film samples have similarly low resistivities, 9.6 µΩ.cm and 0.4 µΩ.cm [3] and 8.7 µΩ.cm and 0.26 µΩ.cm [4] at 300 K and 50 K, respectively. However, in many samples of MgB$_2$ (maybe even a majority) the reported resistivity is much higher. In a polycrystalline bulk sample, Rogado *et al*. report resistivity values of 610 µΩ.cm and 480 µΩ.cm at 300 K and 50 K, respectively [5]. Surprisingly, many high resistivity samples have $T_c$s near 39 K. Perhaps the most unusual results are those of Sharma *et al*. [6], who show that a sample with resistivity at 300 K of above 100 mΩ.cm, well beyond the metal-insulator transition, also has a $T_c$ onset near 39 K.

In addition to these widely variable and sometimes extremely high values of resistivity in samples with $T_c$s of 39 K, Jo *et al*. [7] pointed out a second unusual feature of the transport properties of MgB$_2$. In their "as-made *in situ*" films, they showed that the temperature dependence of the resistivity, from 40 K to 300 K, was identical in form to that observed in single crystals, but increased in magnitude by a factor of 5. They and other authors have suggested that MgB$_2$ samples are perhaps not fully dense. One of the authors (Rowell [8]) has discussed this possibility in more detail, and outlined its implications.

The simplest explanation of the resistivity values of MgB$_2$ is that, in many samples, only a fraction of the cross sectional area of the sample is carrying current. In the case of the films reported by Jo *et al*. [7] the effective cross sectional area is only one fifth of the measured area. This argument can be applied to all MgB$_2$ samples, as long as the temperature dependence of the resistivity, $\rho(T)$, has the same form (roughly $T^2$) as the single crystals. In some bulk samples, the effective area is $10^{-3}$ or less of the measured area. Reduced density (porosity) is not the only means by which the effective area can be reduced. The sample could be fully dense, but the grain boundaries could be largely blocked by MgO, boron oxide, or other impurity phases. A large fraction of the sample could be made of highly resistive or insulating phases (Larbalestier [9] has referred to these phases as "rubble"), rather than superconducting MgB$_2$. Much of the resistivity data, taken at face value, indicates that the individual MgB$_2$ grains are often relatively clean, as $T_c$ is high and $\rho(T)$ has the form seen in single crystals, but connection between these grains is limited. Halbritter *et al*. [10], in a similar explanation of transport in high temperature superconductor materials, have pointed out that meandering percolation paths can increase the effective length of the sample. We have not considered this effect explicitly, but regard it as included in a "decrease in area factor".

An important implication of these high resistivities, if they are indeed due to limitations in the area of the sample that carries current, is that exactly the same limitations should apply to the sample area that carries supercurrent [8]. The $J_c(H = 0)$ of the samples should be inversely proportional to $\rho(300\text{ K}) - \rho(50\text{ K}) \equiv \rho_{300\text{-}50\text{ K}}$. This follows because the ratio of $\Delta\rho$ in fully dense and clean $MgB_2$ to $\Delta\rho$ in the sample is a measure of the reduction in current carrying area. Measurements of $J_c$ by transport are more meaningful in this discussion than those deduced from the Bean model, as current and supercurrent flowing through the whole sample are important.

In this letter, we show that this relationship holds for a number of epitaxial $MgB_2$ films made by hybrid physical-chemical vapor deposition (HPCVD). Details of the *in situ* deposition of $MgB_2$ thin films by HPCVD have been described previously [11]. The samples for this work include films on *c*-cut sapphire, 4H-SiC, and 6H-SiC substrates. The films have *c* axes normal to the substrate surface. Due to large lattice mismatch, the films on sapphire have their hexagonal lattices rotated by 30° to match that of the substrate. No 30° rotation of the hexagonal lattice occurs in films on SiC because of the excellent lattice match of $MgB_2$ with both polytypes of SiC.

Resistivity and $J_c$ measurements were carried out using the standard four-probe technique on patterned bridges. The dimensions of the bridges were either 20 μm (width) × 40 μm (length) or 30 μm (width) × 60 μm (length). They were defined by photo-lithography and ion beam milling. The patterning process was found to cause a slight increase in the residual resistivity and a decrease in $T_c$ of 0.1 - 0.3 K. The thickness of the films ranges from 100 nm to 300 nm, measured using a Dektek profilometer. The product of the bridge width and the film thickness then gives the nominal area used to calculate the resistivity and critical current density from the measured resistance, $R$, and critical current, $I_c$. $I_c$ was determined using a 1-μV criterion from the *I-V* curves.

The $MgB_2$ films deposited by HPCVD consist of hexagonal-shaped growth columns. Fig. 1(a) shows an atomic force microscopy (AFM) image of a $MgB_2$ film on a sapphire substrate. The dimension of the typical growth columns is several hundred nanometers, but gaps can be seen between the growth columns. In Fig. 1(b), a cross-sectional TEM image of another $MgB_2$ film on a sapphire substrate is shown. In this part of the film, the two grains are not in contact at all. While the film as a whole is conducting, in this region neither current nor supercurrent can be carried between these grains (at least in the plane of the picture). The degree of coalescence of the growth columns is influenced by the growth conditions as well as processing conditions such as photo-lithography. In samples where the growth columns are less well connected, the effective area for electrical transport is then smaller than the nominal cross section used for calculating $\rho$ and $J_c$ from the $R$ and $I_c$ measurements.

The resistivity versus temperature dependences for 3 films on three different substrates are shown in Fig. 2. The films in Figs. 2(a) and 2(b) were patterned while that in Fig. 2(c) is unpatterned. It is immediately clear that, although the form of the temperature dependent parts of the resistivities is very similar, the resistivity values differ by a factor of 100. This figure is remarkably similar to Fig. 2 shown in the paper by Putti *et al*. [12], where the resistivities of 3 *bulk* samples of MgB$_2$ are compared. This similarity indicates that the behavior we report here is not unique to thin films, but is also seen in bulk MgB$_2$. The values of $\Delta\rho_{300-50K}$ for the 3 samples are 442, 21.9, and 7.9 μΩ.cm, respectively. This suggests that the effective areas of the higher resistivity samples (Fig. 2(a) and 2(b)) are factors (*F*) of 56 and 2.8 less than the low resistivity sample in Fig. 2(c).

We compare the $J_c$ values in a number of MgB$_2$ films, measured by transport at 5 K and zero magnetic field, in Fig. 3(a), where they are plotted as a function of $\Delta\rho_{300-50K}$ on a log-log scale. These points are from films made as the HPCVD process has been developed over the past 10 months. Two dashed lines with a slope of -1 ($J_c \sim 1/\Delta\rho_{300-50K}$) have been placed through the points showing the maximum (34 MA/cm$^2$) and minimum (0.2 MA/cm$^2$) $J_c$s, which would be the behavior if the reduced effective area *alone* causes the increase of $\Delta\rho_{300-50K}$ and decreases of $J_c$. All the data points fall between the two lines. This implies that although the general trend of $J_c \sim 1/\Delta\rho_{300-50K}$ exists, other effects also influence $J_c$, resulting in a spread of the $J_c$ values between the two lines representing the "reduced effective area" argument alone. For example, many grains in a high resistivity sample could be completely isolated by MgO in the grain boundaries, but a Josephson current might flow between grains where the MgO grain boundary layer is thin. This Josephson current will be smaller than the supercurrent flowing between grains that are well coupled. The resistance of such junctions will also contribute to increasing the resistivity [10]. On the other hand, vortex pinning can be different in different samples. An implication of the data is that maximum pinning (maximum $J_c$) is achieved in these MgB$_2$ films when $\Delta\rho_{300-50K}$ is near 30 μΩ.cm.

The measured residual resistivity at low temperatures, say $\rho(50K)$, is increased not only by the area factor *F*, but also by impurity scattering within the grains, and by the addition of grain boundary resistances, such as the Josephson junctions just discussed. A variation in the intraband π band scattering rate in the two-band picture can also cause the change in the residual resistivity [13]. It is of interest to derive this "intra and inter-grain residual resistivity", $\rho_0$, in the absence of the area effects, i.e., the measured $\rho$ (50 K)/*F*. The result is shown in Fig 3(b). The two unusual data points with low values of $\Delta\rho_{300-50K}$ and $J_c$, but high values of $\rho_0$, are from the very first two superconducting MgB$_2$ films by HPCVD. The remaining films follow the anticipated trend - as $\Delta\rho_{300-50K}$ increases, so does the value of $\rho_0$. However, all the films have $T_c$s above 38 K except for the lowest $J_c$ sample, where $T_c$ = 34.4 K. Hence it seems likely that the increase in $\rho_0$ is primarily due to intergranular effects.

It is interesting to speculate on how universal this behavior might be, across ALL samples of MgB$_2$. If we confine ourselves to those samples which show the roughly $T^2$ dependence of resistivity from 50 K to 300 K, many samples show evidence of having a reduced area to carry current and supercurrent. Unfortunately, only a few papers report both *resistivity* ("resistance" is of no value) and *$J_c$ determined by transport*. However, the literature does seem to indicate the anticipated trends. Samples with low resistivities exhibit $J_c$ of over $10^7$ A/cm$^2$, while samples with higher resistivities have $J_c$s of $10^5$ A/cm$^2$, or even lower values. Our interpretation is that these $J_c$s are reduced primarily because the effective area of the sample is reduced by a factor of 100 or more.

We acknowledge many useful discussions with N. Newman and J. Kim. This work is supported in part by ONR under grant Nos. N00014-02-1-0002 (Rowell), N00014-00-1-0294 (Xi) and N00014-01-1-006 (Redwing), and by NSF under grant Nos. DMR-9876266 and DMR-9972973 (Li), and DMR-9875405 and DMR-9871177 (Pan).

# References

[a] Electronic mail: jmrberkhts@aol.com.

[b] Current address: University of California at Berkeley, Department of Electrical Engineering and Computer Sciences, Berkeley, CA 94720-1770.

[1] Y. Eltsev, K. Nakao, S. Lee, T. Masui, N. Chikumoto, S. Tajima, N. Koshizuka, and M. Murakami, Phys. Rev. B **66**, 180504(R) (2002).

[2] Y. Eltsev, S. Lee, K. Nakao, N. Chikumoto, S. Tajima, N. Koshizuka, and M.Murakami, Physica C **378-381**, 61 (2002).

[3] P. C. Canfield, D. K. Finnemore, S. L. Bud'ko, J. E. Ostenson, G. Lapertot, C. E. Cunningham, and C. Petrovic, Phys. Rev. Lett. **86**, 2423(2001).

[4] A. V. Pogrebnyakov, unpublished.

[5] N. Rogado, M. A. Hayward, K. A. Regan, Y. Wang, N. P. Ong, H. W. Zandbergen, J. M. Rowell, and R. J. Cava, J. Appl. Phys. **91**, 274 (2002).

[6] P. A. Sharma, N. Hur, Y. Horibe, C. H. Chen, B. G. Kim, S. Guha, M. Z. Cieplak, and S.-W. Cheong, Phys. Rev. Lett. **89**, 167003 (2002).

[7] W. Jo, J.-U. Huh, T. Ohnishi, A. F. Marshall, M. R. Beasley, and R. H. Hammond, Appl. Phys. Lett. **80,** 3563 (2002).

[8] J. M. Rowell, to be published in Supercond. Sci. Tech.

[9] D. C. Larbalestier, private communications.

[10] J. Halbritter, M. R. Dietrich, H. Kupfer, B. Runtsch, J. Wuhl, Z. Phys. B **71**, 411 (1988).

[11] X. H. Zeng, A. V. Pogrebnyakov, A. Kotcharov, J. E. Jones, X. X. Xi, E. M. Lysczek, J. M. Redwing, S. Y. Xu, Q. Li, J. Lettieri et al., Nature Materials **1**, 35 (2002).

[12] M. Putti, V. Braccini, E. Galleani, F. Napoli, I. Pallecchi, A. S. Siri, P. Manfrinetti, and A. Palenzona, cond-mat/0210047 (2002).

[13] I. I. Mazin, O. K. Andersen, O. Jepsen, O. V. Dolgov, J. Kortus, A. A. Golubov, A. B. Kuz'menko, and D. van der Marel, Phys. Rev. Lett. **89**, 107002 (2002).

# Figure Captions

Fig. 1: (a) AFM image of an epitaxial $MgB_2$ film grown by HPCVD on a sapphire substrate. (b) Cross-section TEM image of a $MgB_2$ film on sapphire substrate. In this image, two growth columns are shown to be not in contact.

Fig. 2: The temperature dependence of resistivity for 3 $MgB_2$ films on different substrates. Films in (a) and (b) were patterned while that in (c) was unpatterned.

Fig. 3: (a) $J_c$ ($T = 5$ K, $H = 0$) and (b) $\rho_0 \equiv \rho$ (50 K)$/F$ as a function of $\Delta\rho_{300\text{-}50K} \equiv \rho$ (300 K) - $\rho$ (50 K) for a collection of $MgB_2$ films on different substrates. The variation in $\Delta\rho_{300\text{-}50K}$ is likely due to the different processing conditions of the films. The dashed lines represent $J_c \sim 1/\Delta\rho_{300\text{-}50K}$.

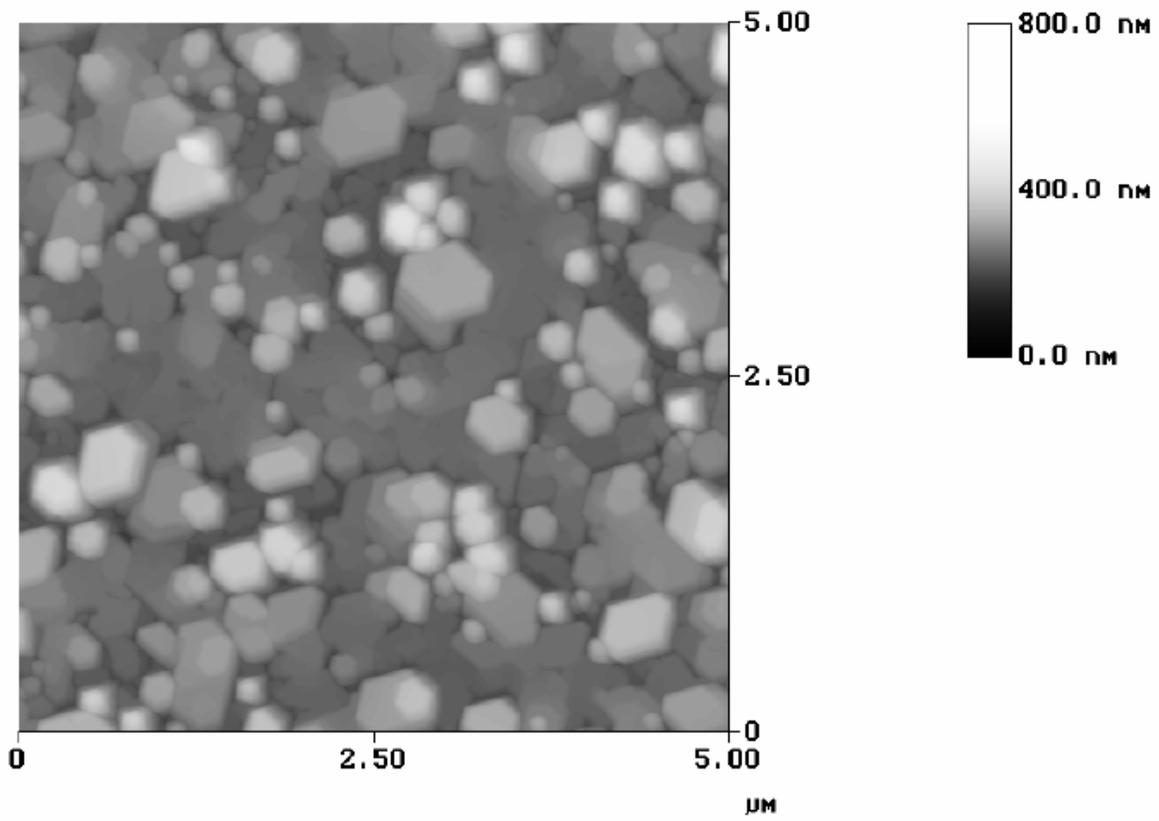

Fig. 1(a) of 3
Rowell et al.

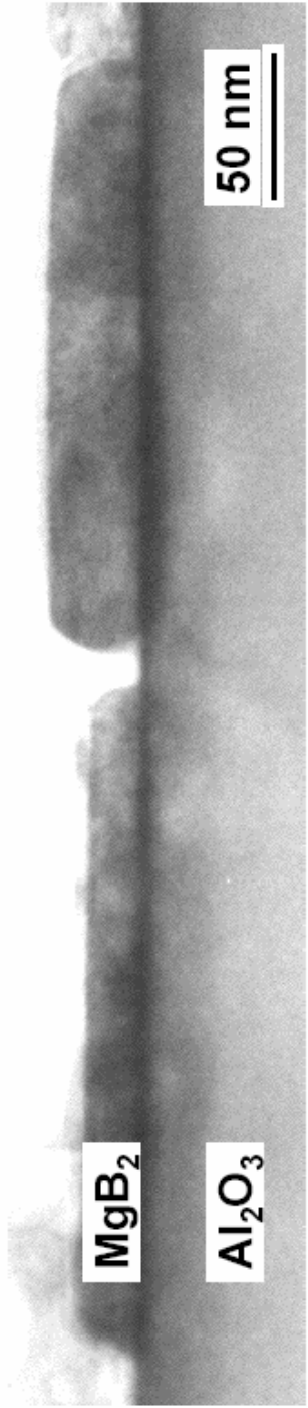

Fig. 1(b) of 3
Rowell et al.

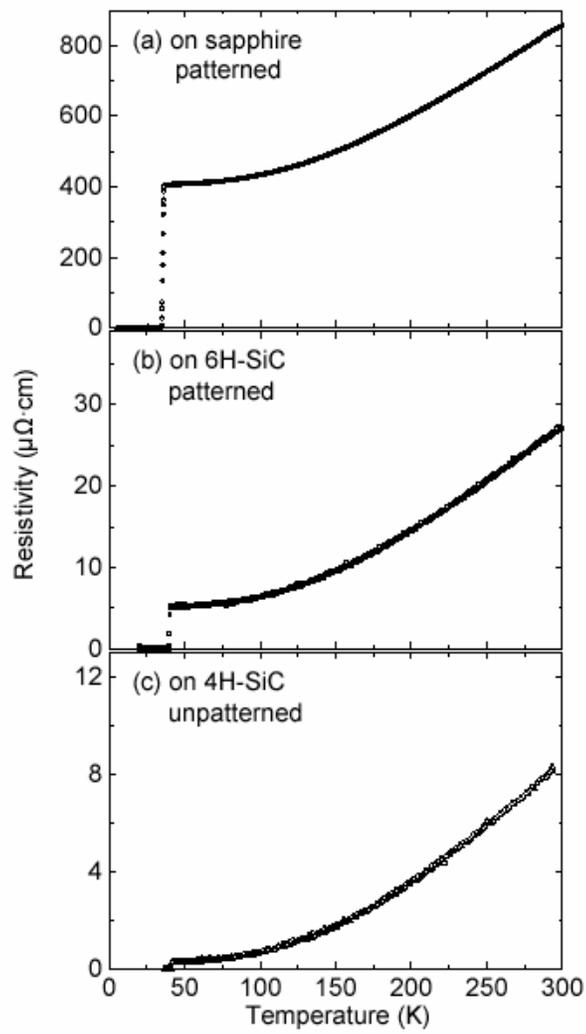

**Fig. 2 of 3
Rowell et al.**

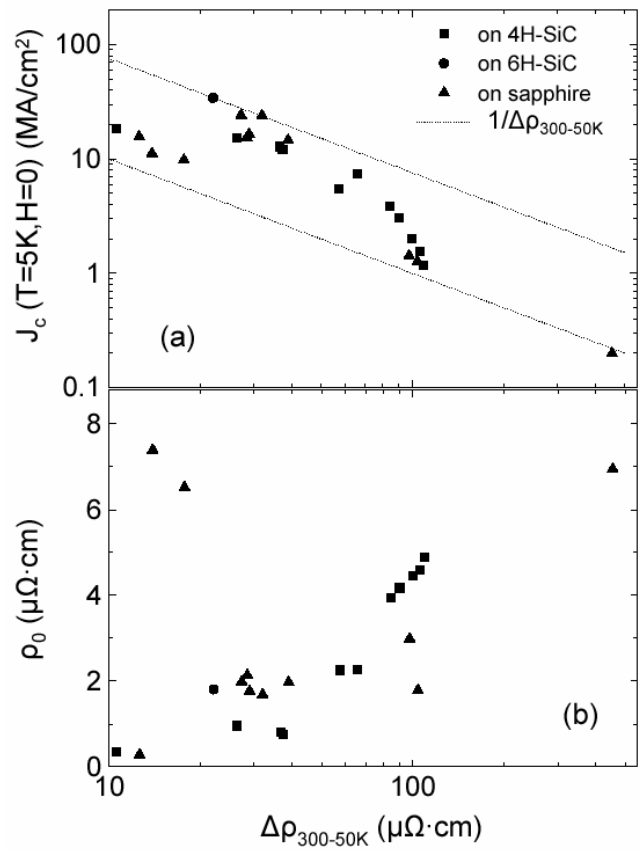

**Fig. 3 of 3
Rowell et al.**